\pdfoutput=1
\documentclass[11pt, reqno]{article}
\usepackage{jheplike}
\usepackage{graphicx}
\usepackage{amssymb}
\usepackage{amsmath}
\usepackage{mathrsfs}
\usepackage{hyperref}
\usepackage{multirow}
\usepackage[compat=1.1.0]{tikz-feynman}
\usepackage[utf8]{inputenc}

\newcommand{\be}{\begin{equation}}
\newcommand{\ee}{\end{equation}}

\newcommand{\Li}{\textrm{Li}}

\newcommand{\EE}[3]{\mathcal{E}_4 \left( \begin{matrix} #1 \\ #2 \end{matrix} \, ; #3 , \vec{a}\right)}

\newcommand{\Gt}[3]{\tilde{\Gamma}\left( \begin{matrix} #1 \\ #2 \end{matrix} \, ; #3 ,\tau\right)}

\title{A Three-Parameter Elliptic Double-Box}

\author{Alex Chaparro Pozo,}
\author{Matt von Hippel,}

\affiliation{Niels Bohr International Academy, Niels Bohr Institute, University of Copenhagen, Blegdamsvej 17,
2100 Copenhagen, Denmark}

\abstract{We express a toy model of the ten-point elliptic double-box, first characterized in arXiv:1712.02785, in terms of elliptic polylogarithms. This toy model corresponds to a particular unphysical limit of the elliptic double-box in which it depends on only three dual conformal cross-ratios. While the diagram is fully permutation symmetric in the cross-ratios in this limit, this property is not manifest in either of the two elliptic polylogarithm formalisms we use to express it. We observe that the function is a pure elliptic polylogarithm, which is the result of nontrivial identities between elliptic integrals depending on the conformal cross-ratios.}

\emailAdd{mvonhippel@nbi.ku.dk}

\begin{document}
\hypersetup{pageanchor=false}

\maketitle
\hypersetup{pageanchor=true}

\section{Introduction}

Recent years have seen enormous progress in tackling integrals beyond polylogarithms in perturbative quantum field theory, particularly those involving elliptic curves. The first such integral to be considered was a two-loop propagator correction of massive scalars in two dimensions called the sunrise integral. Due to its comparatively simple and self-contained nature, this integral has been the focus of a wide array of investigations, used as a testbed for new formalisms and new insights~\cite{SABRY1962401,Broadhurst:1993mw,Berends:1993ee,Bauberger:1994by,Bauberger:1994hx,Bauberger:1994nk,Caffo:1998du,Laporta:2004rb,Muller-Stach:2011qkg,Adams:2013nia,Adams:2013kgc,Remiddi:2013joa,Bloch:2013tra,Adams:2014vja,Adams:2015gva,Adams:2015ydq,Remiddi:2016gno,Bloch:2016izu,Adams:2017ejb,Broedel:2018iwv,Broedel:2017siw,Bourjaily:2018yfy,Bogner:2019lfa,Campert:2020yur,Frellesvig:2021vdl}. One result of this work has been the development of a powerful and versatile formalism for these integrals, one which can convert between a representation as iterated integrals on a toroidal domain~\cite{brown2011multiple,Broedel:2017kkb} and one as iterated integrals involving roots of quartic polynomials, normalized to preserve a certain notion of ``purity''~\cite{Broedel:2018qkq}.

The first massless Feynman integral involving an elliptic curve is more complicated, a ten-particle two-loop double-box integral that gives a particular supercomponent of the corresponding amplitude in planar $\mathcal{N}=4$ super Yang-Mills theory~\cite{Caron-Huot:2012awx,Bourjaily:2017bsb}. This integral has recently been expressed in terms of the above formalism~\cite{Kristensson:2021ani}, leading to proposals for extending that formalism~\cite{Wilhelm:2022wow}. While there has been a great deal of progress with this integral, its kinematic complexity makes it challenging to work with.

In ref.~\cite{Bourjaily:2017bsb}, one of the authors described a limit of the ten-particle double-box that preserves its ellipticity. This limit is a toy model with no particular physical relevance, in which certain pairs of non-adjacent dual points are taken to be light-like separated. This toy model has the advantage that the double-box becomes dependent on only three parameters, with full permutation symmetry in those parameters. Ref.~\cite{Bourjaily:2017bsb} represented this integral in terms of a one-fold integral over weight-three polylogarithms, and observed that in this form its permutation symmetry is not manifest.

We believe that this toy model could be a useful tool for investigation of elliptic Feynman integrals in future. It is intermediate in complexity between the sunrise integral and the full ten-point double-box, and the challenge of making its symmetry manifest may suggest improvements to the formalism for elliptic Feynman integrals. To explore the toy model double-box further, we express it in modern formalism: iterated integrals on a torus (known as $\tilde{\Gamma}$ functions) and iterated integrals involving roots of quartic polynomials (specifically the pure basis denoted by $\mathcal{E}_4$). We find that in either formalism the symmetries of the diagram are still not manifest.

We organize the rest of the paper as follows. In section~\ref{sec:toydef} we give a definition of the toy model we will consider. Section~\ref{sec:empls} reviews the bases of elliptic multiple polylogarithm functions we will be using. We describe our procedure to convert the results of ref.~\cite{Bourjaily:2017bsb} into this formalism in section~\ref{sec:procedure}, and present our results in section~\ref{sec:results}. Finally, we conclude and mention some open questions.

\section{The Toy Model}
\label{sec:toydef}

\begin{figure}
     \centering
     \includegraphics[width=0.5\textwidth]{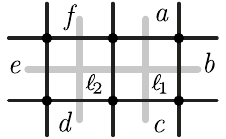}
     \caption{The elliptic double-box. Labels denote dual points $x_i$.}
     \label{fig:graph}
\end{figure}

In planar $\mathcal{N}=4$ super Yang-Mills, the leading contribution to a particular supercomponent of the ten-particle amplitude consists of one two-loop graph, a double-box depicted in figure~\ref{fig:graph}.

For this ten-particle amplitude all external legs are massless. We can define a dual position space with $p_i=x_{i+1}-x_i$. This diagram is finite, and as such is invariant under a conformal symmetry in the dual space, or dual conformal symmetry~\cite{Drummond:2006rz, Bern:2006ew, Bern:2007ct, Alday:2007hr, Bern:2008ap, Drummond:2008vq}. In terms of the dual points, masslessness of the middle legs of the double-box implies that the dual points $f$ and $a$ are light-like separated, as well as the dual points $d$ and $c$. Writing $(a,b)\equiv (x_a-x_b)^2\equiv x_{ab}^2$, we have,
\be
(a,f)=(d,c)=0\,.
\ee
Following ref.~\cite{Bourjaily:2017bsb}, we impose additional restrictions on the kinematics:
\be
(f,b)=(b,d)=(c,e)=(e,a)=0\,.
\ee
One can think of these restrictions as letting us reduce to the kinematics of six massless particles, with $\{x_a,x_b,x_c,x_d,x_e,x_f\}$ becoming $\{x_1,x_3,x_5,x_4,x_6,x_2\}$. However, as these dual points are in a different order than our original dual points, this is not a sensible massless limit of the diagram itself. It is for this reason that we refer to this limit as a toy model.

In this limit, the diagram depends on three dual-conformal cross-ratios,
\be
u_1\equiv\frac{(a,b)(d,e)}{(a,d)(b,e)}\,,\quad u_2\equiv\frac{(f,d)(c,a)}{(f,c)(d,a)}\,,\quad u_3\equiv\frac{(b,c)(e,f)}{(b,e)(c,f)}\,.
\ee

An alternative presentation for these cross-ratios, in terms of so-called $y$ variables, will be useful later, as it rationalizes a square root that would otherwise frequently appear in this calculation. The cross-ratios $u_i$ are given in terms of these variables as follows:
\begin{align}
u_1=\frac{y_1(1-y_2)(1-y_3)}{(1-y_1 y_2)(1-y_3 y_1)}\nonumber\\
u_2=\frac{y_2(1-y_3)(1-y_1)}{(1-y_2 y_3)(1-y_1 y_2)}\nonumber\\
u_3=\frac{y_3(1-y_1)(1-y_2)}{(1-y_3 y_1)(1-y_2 y_3)}
\label{eq:yvars}
\end{align}

Ref.~\cite{Bourjaily:2017bsb} described how this diagram can be Feynman-parametrized loop-by-loop in a way that preserves its dual conformal symmetry. This resulted in an expression of the form,
\be
I_{\textrm{toy}}^{\textrm{ell}}=\int^\infty_0 d\alpha \int^\infty_0 d^3\vec{\beta} \frac{1}{f_1 f_2 f_3}\,,
\ee
where we will for now neglect the (Feynman-parameter-independent) numerator. The polynomials $f_1, f_2, f_3$ are defined as,
\begin{align}
f_1&=\beta_1+\beta_2+\beta_1\beta_2\\
f_2&=1+\alpha u_1+u_2 \beta_3\\
f_3&=f_1+\alpha(\beta_1+u_3\beta_3)+\beta_2\beta_3\,.
\end{align}

In the above reference, the authors were able to integrate in the three $\beta_i$ parameters, yielding an expression for the toy model in terms of an integral over weight-three polylogarithmic functions, with a prefactor involving a square root of a quartic equation in the remaining integration variable:
\be
I_{\textrm{toy}}^{\textrm{ell}}=\int^\infty_0 d\alpha \frac{1}{\sqrt{Q(\alpha)}}H_{\textrm{toy}}(\alpha)\,,
\label{eq:initial}
\ee
where,
\be
Q(\alpha)\equiv (1+\alpha(u_1+u_2+u_3+\alpha u_1 u_3))^2-4\alpha(1+\alpha u_1)^2 u_3\,,
\ee
and $H_{\textrm{toy}}=F_1-F_2$, with
\begin{align}
F_i(\alpha)\equiv & G(\bar{w}_i,0,0;\alpha)+G(\bar{w}_i,\bar{0},\bar{0};\alpha)-G(\bar{w}_i,0,\bar{0};\alpha)
-G(\bar{w}_i,\bar{0},0;\alpha)\nonumber\\&-G(\bar{w}_i,-\bar{w}_1\bar{w}_2,0;\alpha)-G(\bar{w}_i,\frac{\bar{w}_1\bar{w}_2}{\overline{w_1+w_2}},\bar{0};\alpha)
+G(\bar{w}_i,\frac{\bar{w}_1\bar{w}_2}{\overline{w_1+w_2}};\alpha)\log(w_1w_2\bar{w}_1\bar{w}_2)
\nonumber\\&-G(\bar{w}_i,-\bar{w}_1\bar{w}_2;\alpha)\log\left(\frac{-1}{\bar{w}_1\bar{w}_2}\right)
+\Big\{G(\bar{w}_i,0;\alpha)-G(\bar{w}_i,\bar{0};\alpha)\Big\}\log(-w_1w_2)
\nonumber\\&+G(\bar{w}_i;\alpha)\Big\{\frac{1}{2}\log^2\left(\frac{1}{\overline{w_1+w_2}}\right)+\log(w_1w_2)
\log\left(\frac{-1}{\bar{w}_1\bar{w}_2}\right)\nonumber\\&
\quad\quad\quad\quad\quad-\log(\frac{1}{\overline{w_1+w_2}})\log(\frac{1}{\bar{w}_1\bar{w}_2})+
\Li_2(-\frac{\overline{w_1+w_2}}{\bar{w}_1\bar{w}_2})\Big\}\,,
\end{align}
where we use the compressed notation $\bar{x}\equiv-1/(1+x)$ (so that $\bar{0}=-1$), and
\be
w_{1,2}\equiv\left[\alpha((\alpha u_3-1)u_1-u_2+u_3)-1\pm\sqrt{Q(\alpha)}\right]/(2\alpha u_2)\,.
\ee
Here, the functions $G$ are the usual multiple polylogarithms, with conventions,
\be
G(a_1,\ldots,a_n;z)=\int_0^z \frac{dt}{t-a_1}G(a_2,\ldots,a_n;t)\,.
\ee

This expression for the toy model double-box will be the starting point for our analysis here.

\section{Elliptic Multiple Polylogarithms}
\label{sec:empls}

Several distinct formalisms for elliptic polylogarithms exist in the literature. There are two we will make use of in this paper. One, the pure elliptic multiple polylogarithms of ref.~\cite{Broedel:2018qkq}, is for us a more useful starting-point, as it is easier to obtain from our initial expression in eq.~\ref{eq:initial}. The other, the meromorphic iterated integrals on a torus described in~\cite{brown2011multiple,Broedel:2017kkb}, are more difficult to work with, but they do have one property that will be important for us: they can be made to depend on a basis of functions that is closed under permutations of the cross-ratios $u_i$. As such, we will here review both formalisms.

\subsection{Pure Elliptic Multiple Polylogarithms}
\label{subsec:pureempls}

The pure elliptic multiple polylogarithms of ref.~\cite{Broedel:2018qkq} are iterated integrals with integration kernels that depend on the square root of a cubic or quartic polynomial. There are many possible bases of such integration kernels, but this particular formalism aims for a basis such that the integrals are \textit{pure}: functions that are unipotent (satisfying a differential equation without a homogeneous term), and with total differential containing only pure functions and one-forms with logarithmic singularities. These are the functions expected to be most directly analogous to polylogarithms, and, with appropriate normalization, to constitute the best bases for scattering amplitudes.

Here we will focus on the version of this formalism defined using the square root of a quartic polynomial\footnote{There are subtle differences for a cubic polynomial that will not be relevant here.}. These functions are iterated integrals on a set of integration kernels $\Psi_n(c,x,\vec{a})$, where $n\in\mathbb{Z}$, $c$ is a branch point which can be $\infty$, $x$ is the integration variable, and $\vec{a}$ is a list of the roots of the polynomial in a particular ordering. These functions are denoted as follows:
\be
\EE{n_1\ldots n_k}{c_1\ldots c_k}{x}=\int_0^x dt\, \Psi_{n_1}(c_1,t,\vec{a})\,\EE{n_2\ldots n_k}{c_2\ldots c_k}{t}\,.
\ee
The presence of the list $\vec{a}$ (and its ordering) in the formalism is what will make the basis of functions that results not closed under permutation of the cross-ratios of our problem.

In practice, we need only consider $\Psi_n(c,x,\vec{a})$ for values of $n$ close to zero. In our case, we need only the following:
\begin{subequations}
\label{eq:purekernels}
\begin{align}
\Psi_0(0,x,\vec{a})&=\frac{c_4}{\omega_1 y}\label{eq:kernel0}\\
\Psi_1(c,x,\vec{a})&=\frac{1}{x-c}\label{eq:kernel1c}\\
\Psi_{-1}(c,x,\vec{a})&=\frac{y_c}{y(x-c)}+Z_4(c,\vec{a})\frac{c_4}{y}\label{eq:kernelm1c}\\
\Psi_1(\infty,x,\vec{a})&=-Z_4(x,\vec{a})\frac{c_4}{y}\label{eq:kernel1inf}\\
\Psi_{-1}(\infty,x,\vec{a})&=\frac{x}{y}-\frac{1}{y}\left[a_1+2c_4 G_*(\vec{a})\right]\,,\label{eq:kernelm1inf}
\end{align}
\end{subequations}
where we have,
\be
y^2=P_4(x)=(x-a_1)(x-a_2)(x-a_3)(x-a_4)\,,
\ee
so that $y$ is the square root of the quartic polynomial that defines our elliptic curve (normalized so the coefficient of $x^4$ is one), while,
\be
c_4\equiv\frac{1}{2}\sqrt{(a_1-a_3)(a_2-a_4)}\,.
\ee

An elliptic curve has two periods, for which we will use,
\begin{align}
\omega_1&=2c_4\int_{a_2}^{a_3}\frac{dx}{y}\nonumber\\
\omega_1&=2c_4\int_{a_1}^{a_2}\frac{dx}{y}\,,
\end{align}
and two quasi-periods,
\begin{align}
\eta_1&=-\frac{1}{2}\int_{a_2}^{a_3}\frac{dx}{c_4 y}\left[x^2-\frac{s_1(\vec{a})}{2}x+\frac{s_2(\vec{a})}{6}\right]\nonumber\\
\eta_2&=-\frac{1}{2}\int_{a_1}^{a_2}\frac{dx}{c_4 y}\left[x^2-\frac{s_1(\vec{a})}{2}x+\frac{s_2(\vec{a})}{6}\right]\nonumber\\
\end{align}

The function $Z_4(x,\vec{a})$ is designed to have a simple pole at infinity. It is defined by the following integral,
\be
Z_4(x,\vec{a})\equiv\int_{a_1}^x \frac{dx'}{y} \left[\frac{1}{c_4}\left(x^2-\frac{s_1(\vec{a})}{2}x+\frac{s_2(\vec{a})}{6}\right) +4c_4\frac{\eta_1}{\omega_1}\right]\,,
\ee
where $s_1$ and $s_2$ are symmetric polynomials. 

Finally, we have
\be
G_*(\vec{a})\equiv\frac{1}{\omega_1}g^{1}(z_*,\tau)\,,
\ee
where
\be
z_*\equiv\frac{c_4}{\omega_1}\int_{a_1}^{-\infty}\frac{dx}{y}\,,
\ee
and $g^{(1)}$ and $\tau$ will be defined in the next subsection.

\subsection{\texorpdfstring{$\tilde{\Gamma}$}{Gammatilde} Formalism}
\label{subsec:gammatilde}

It is well-known that an elliptic curve defined over the complex numbers is isomorphic to a complex torus. The modular parameter, $\tau$, of this torus can be defined in terms of the periods of the elliptic curve like so:
\be
\tau=\frac{\omega_2}{\omega_1}\,.
\ee

Given a point $(x,y)$ on the elliptic curve we can find a corresponding point on the torus using Abel's map as follows,
\be
z_x=\frac{c_4}{\omega_1}\int_{a_1}^{x}\frac{dx'}{y}\,.
\ee
Referencing the previous section it should immediately become clear that $z_*$ is the image of $-\infty$ under this map.

With these definitions, we can state the formalism for elliptic multiple polylogarithms as iterated integrals on a torus, as a class of functions $\tilde{\Gamma}$:
\be
\Gt{n_1 & \ldots & n_k}{z_1 & \ldots & z_k}{z}=\int_0^z dz' g^{(n_1)}(z'-z_1,\tau)\Gt{n_2 & \ldots & n_k}{z_2 & \ldots & z_k}{z}\,,
\ee
where the functions $g^{(n)}(z,\tau)$ are defined via the Eisenstein-Kronecker series,
\be
F(z,\alpha,\tau)=\frac{1}{\alpha}\sum_{n\geq 0}g^{(n)}(z,\tau)\alpha^n=\frac{\theta'_1(0,\tau)\theta_1(z+\alpha,\tau)}{\theta_1(z,\tau)\theta_1(\alpha,\tau)}\,,
\ee
where $\theta_1$ is the odd Jacobi theta function and $\theta'_1$ is its derivative with respect to its first argument.

These functions are related to the functions in the previous section in a very simple way,
\begin{align}
dx \Psi_{\pm n}(c,x,\vec{a})& = \nonumber\\
dz_x\Big[g^{(n)} & (z_x-z_c,\tau)\pm g^{(n)}(z_x+z_c,\tau)-\delta_{\pm n,1}\left(g^{(1)}(z_x-z_*,\tau)+g^{(1)}(z_x+z_*,\tau)\right)\Big]\,.
\label{eq:kernelrelation}
\end{align}

\section{Procedure}
\label{sec:procedure}

In principle, the procedure to re-express an expression of the form~\ref{eq:initial} in elliptic polylogarithms was described in full in refs.~\cite{Broedel:2017kkb,Broedel:2017siw}. While our situation is slightly different in that we integrate directly into the pure elliptic polylogarithms of ref.~\cite{Broedel:2018qkq}, the methodology is similar enough that we will not repeat the details of those references here. Instead, we will sketch the method, and go into detail only for those aspects which are particular subtleties of our situation.

To begin, we prefer to avoid integrating from zero to infinity as this will result in functions that are difficult to check numerically, so we map the integration domain of eq.~\ref{eq:initial} to an integral from zero to one by mapping to a new variable,
\be
x=\frac{\alpha}{1+\alpha}\,.
\ee

In general, the method of refs.~\cite{Broedel:2017kkb,Broedel:2017siw} demands that we re-write the multiple polylogarithms in the integrand of~\ref{eq:initial} in terms of (pure) elliptic multiple polylogarithms, with their only dependence on the integration variable residing in their function argument (and thus the endpoint of integration). We must do this recursively by ``re-fibering'' the functions in the integrand, taking derivatives\footnote{To perform these derivatives and otherwise manipulate multiple polylogarithms, we made use of the package PolyLogTools~\cite{Duhr:2019tlz}.} until we reach algebraic functions and then integrating them back up, using partial-fractioning and integration-by-parts to associate them with the integration kernels in eq.~\ref{eq:purekernels}. We can then integrate our function back into pure elliptic multiple polylogarithms and perform a numerical sanity check. 

In general, this procedure will not result in manifestly pure functions: instead, there will be pure elliptic multiple polylogarithms with various coefficients, including rational and algebraic functions in the $y_i$ and the function $Z_4$ with various arguments. However, we do expect every multiple polylogarithm (depending on an elliptic curve) to be representable in terms of pure elliptic multiple polylogarithms. In order for this to be the case, certain identities have to hold true, such that certain linear combinations of functions $Z_4$ are algebraic in the $y_i$. Using this expectation we have identified several candidate identities of this form and verified them numerically:
\begin{align}
\label{eq:id1}
&Z_4(1,\vec{a})= -\frac{A}{2 c_4 q_4 (1 - y_1 y_2) y_3 (1 - y_1 y_3)^2 (1 -
          y_2 y_3)} \nonumber\\
          &\quad\quad\quad\quad\quad\quad\quad\quad\quad\quad\quad\quad\quad\quad\quad+ \frac{1}{2} Z_4\left(\frac{(1 - y_1 y_3)}{(y_3 - y_1 y_3)},\vec{a}\right) +
   \frac{1}{2} Z_4\left(\frac{(1 - y_1 y_3)}{(1 - y_1)},\vec{a}\right)\\
   \label{eq:id2}
 &Z_4(0,\vec{a}) =  -\frac{B}{6 c_4 q_4 (1 - y_1 y_2) y_3 (1 - y_1 y_3) (1 -
          y_2 y_3)} \nonumber\\
          &\quad\quad\quad\quad\quad\quad\quad\quad\quad\quad\quad\quad\quad\quad\quad+ \frac{1}{2} Z_4\left(\frac{(1 - y_1 y_3)}{(y_3 - y_1 y_3)},\vec{a}\right) +
   \frac{1}{2} Z_4\left(\frac{(1 - y_1 y_3)}{(1 - y_1)},\vec{a}\right)\\
   \label{eq:id3}
 &Z_4\left(\frac{(1 - y_1 y_2) (1 - y_1 y_3)}{(1 - y_1) (1 -
      y_1 y_2 y_3)},\vec{a}\right) = -\frac{C}{6 c_4 q_4 (1 - y_1 y_2) y_3 (1 - y_1 y_3) (1 -   y_2 y_3) (1 - y_1 y_2 y_3)} \nonumber\\
          &\quad\quad\quad\quad\quad\quad\quad\quad\quad\quad\quad\quad\quad\quad\quad+
   \frac{1}{2} Z_4\left(\frac{(1 - y_1 y_3)}{(y_3 - y_1 y_3)},\vec{a}\right) + \frac{1}{2} Z_4\left(\frac{(1 - y_1 y_3)}{(1 - y_1)},\vec{a}\right)
\end{align}
where,
\be
q_4=\sqrt{\frac{(1 - y_1)^2 (1 - y_2)D}{(1 - y_1 y_2)^2 (1 -
       y_1 y_3)^4 (1 - y_2 y_3)^2)} }\,,
       \label{eq:q4def}
\ee
and $A, B, C$, and $D$ are polynomials,
\begin{align}
A=&1 - y_2 - 3 y_1 y_3 + 2 y_2 y_3 + y_1 y_2 y_3 + y_3^2 -
        2 y_1 y_3^2 + 4 y_1^2 y_3^2 - 3 y_2 y_3^2 + 2 y_1 y_2 y_3^2 -
        y_1^2 y_2 y_3^2 \nonumber\\
        &- 2 y_1 y_2^2 y_3^2 + y_1^2 y_2^2 y_3^2 + y_1 y_3^3 -
        2 y_1^2 y_3^3 - y_1 y_2 y_3^3 + 2 y_1^2 y_2 y_3^3 - 3 y_1^3 y_2 y_3^3 +
        4 y_1 y_2^2 y_3^3 - 2 y_1^2 y_2^2 y_3^3 \nonumber\\
        &+ y_1^3 y_2^2 y_3^3 +
        y_1^2 y_2 y_3^4 + 2 y_1^3 y_2 y_3^4 - 3 y_1^2 y_2^2 y_3^4 -
        y_1^3 y_2 y_3^5 +
        y_1^3 y_2^2 y_3^5\nonumber\\
B=&3 - 3 y_2 + 6 y_3 - 8 y_1 y_3 + 4 y_2 y_3 - 2 y_1 y_2 y_3 +
        y_3^2 - 2 y_1 y_3^2 - 11 y_2 y_3^2 + 11 y_1^2 y_2 y_3^2 +
        2 y_1 y_2^2 y_3^2 \nonumber\\
        &- y_1^2 y_2^2 y_3^2 + 2 y_1 y_2 y_3^3 -
        4 y_1^2 y_2 y_3^3 + 8 y_1 y_2^2 y_3^3 - 6 y_1^2 y_2^2 y_3^3 +
        3 y_1^2 y_2 y_3^4 -
        3 y_1^2 y_2^2 y_3^4\nonumber\\
C=&3 - 3 y_2 - 4 y_1 y_3 + 2 y_2 y_3 - 7 y_1 y_2 y_3 +
        9 y_1 y_2^2 y_3 - y_3^2 + 2 y_1 y_3^2 - y_2 y_3^2 + 12 y_1 y_2 y_3^2 +
        11 y_1^2 y_2 y_3^2 \nonumber\\
        &- 16 y_1 y_2^2 y_3^2 - 7 y_1^2 y_2^2 y_3^2 -
        7 y_1 y_2 y_3^3 - 16 y_1^2 y_2 y_3^3 + 11 y_1 y_2^2 y_3^3 +
        12 y_1^2 y_2^2 y_3^3 - y_1^3 y_2^2 y_3^3 + 2 y_1^2 y_2^3 y_3^3 \nonumber\\
        &-
        y_1^3 y_2^3 y_3^3 + 9 y_1^2 y_2 y_3^4 - 7 y_1^2 y_2^2 y_3^4 +
        2 y_1^3 y_2^2 y_3^4 - 4 y_1^2 y_2^3 y_3^4 - 3 y_1^3 y_2^2 y_3^5 +
        3 y_1^3 y_2^3 y_3^5\nonumber\\
D=&1 - y_2 + 2 y_3 - 4 y_1 y_3 + 2 y_2 y_3 +
       y_3^2 - 4 y_1 y_3^2 + 4 y_1^2 y_3^2 - 5 y_2 y_3^2 - 4 y_1 y_2 y_3^2 +
       6 y_1^2 y_2 y_3^2 \nonumber\\
       &+ 2 y_1^2 y_2^2 y_3^2 + 4 y_1 y_2 y_3^3 +
       8 y_1^2 y_2 y_3^3 - 8 y_1^3 y_2 y_3^3 + 8 y_1 y_2^2 y_3^3 -
       8 y_1^2 y_2^2 y_3^3 - 4 y_1^3 y_2^2 y_3^3 - 2 y_1^2 y_2 y_3^4 \nonumber\\
       &-
       6 y_1^2 y_2^2 y_3^4 + 4 y_1^3 y_2^2 y_3^4 + 5 y_1^4 y_2^2 y_3^4 -
       4 y_1^2 y_2^3 y_3^4 + 4 y_1^3 y_2^3 y_3^4 - y_1^4 y_2^3 y_3^4 -
       2 y_1^4 y_2^2 y_3^5 + 4 y_1^3 y_2^3 y_3^5 \nonumber\\
       &- 2 y_1^4 y_2^3 y_3^5 +
       y_1^4 y_2^2 y_3^6 - y_1^4 y_2^3 y_3^6\,.        
\end{align}

We do not have analytic proof of these identities, but we expect this to be possible to find using Abel's theorem, along the lines described in ref.~\cite{Wilhelm:2022wow}.

Once we have an expression in terms of pure elliptic multiple polylogarithms, it is straightforward to transform our expression to one involving the $\tilde{\Gamma}$ basis using eq.~\ref{eq:kernelrelation}.

\section{Results}
\label{sec:results}

Performing the procedure in the above section, we find,
\be
I_{\textrm{toy}}^{\textrm{ell}}=\frac{\omega_1}{c_4 q_4}T\,,
\ee
where $q_4$ is defined in~\ref{eq:q4def} and,
{\allowdisplaybreaks
\begin{align}
T&=3 \EE{0&-1&1&1}{0&0&0&1}{1}-2 \EE{0&-1&1&1}{0&0&1&0}{1}+3 \EE{0&-1&1&1}{0&0&1&1}{1}\nonumber\\
&-\EE{0&-1&1&1}{0&0&1&b_1}{1}-3 \EE{0&-1&1&1}{0&0&b_2&0}{1}-2 \EE{0&-1&1&1}{0&0&b_1&1}{1}\nonumber\\
&+\EE{0&-1&1&1}{0&0&b_3&1}{1}-\EE{0&-1&1&1}{0&0&b_3&b_1}{1}+3 \EE{0&-1&1&1}{0&0&b_4&0}{1}\nonumber\\
&-3 \EE{0&-1&1&1}{0&0&b_4&1}{1}-\EE{0&-1&1&1}{0&1&0&1}{1}-2 \EE{0&-1&1&1}{0&1&0&b_1}{1}\nonumber\\
&+3 \EE{0&-1&1&1}{0&1&1&0}{1}-3 \EE{0&-1&1&1}{0&1&1&1}{1}+2 \EE{0&-1&1&1}{0&1&1&b_1}{1}\nonumber\\
&+\EE{0&-1&1&1}{0&1&b_2&0}{1}-2 \EE{0&-1&1&1}{0&1&b_1&0}{1}+3 \EE{0&-1&1&1}{0&1&b_1&1}{1}\nonumber\\
&-2 \EE{0&-1&1&1}{0&1&b_1&b_1}{1}-\EE{0&-1&1&1}{0&1&b_3&1}{1}+\EE{0&-1&1&1}{0&1&b_3&b_1}{1}\nonumber\\
&-3 \EE{0&-1&1&1}{0&1&b_4&0}{1}+3 \EE{0&-1&1&1}{0&1&b_4&1}{1}-\EE{0&-1&1&1}{0&b_5&0&1}{1}\nonumber\\
&+\EE{0&-1&1&1}{0&b_5&0&b_1}{1}-\EE{0&-1&1&1}{0&b_5&1&0}{1}+\EE{0&-1&1&1}{0&b_5&b_2&0}{1}\nonumber\\
&+\EE{0&-1&1&1}{0&b_5&b_1&0}{1}-\EE{0&-1&1&1}{0&b_5&b_1&1}{1}+\EE{0&-1&1&1}{0&b_5&b_1&b_1}{1}\nonumber\\
&+\EE{0&-1&1&1}{0&b_5&b_3&1}{1}-\EE{0&-1&1&1}{0&b_5&b_3&b_1}{1}-\EE{0&-1&1&1}{0&b_6&0&1}{1}\nonumber\\
&+\EE{0&-1&1&1}{0&b_6&0&b_1}{1}-\EE{0&-1&1&1}{0&b_6&1&0}{1}+\EE{0&-1&1&1}{0&b_6&b_2&0}{1}\nonumber\\
&+\EE{0&-1&1&1}{0&b_6&b_1&0}{1}-\EE{0&-1&1&1}{0&b_6&b_1&1}{1}+\EE{0&-1&1&1}{0&b_6&b_1&b_1}{1}\nonumber\\
&+\EE{0&-1&1&1}{0&b_6&b_3&1}{1}-\EE{0&-1&1&1}{0&b_6&b_3&b_1}{1}+\EE{0&-1&1&1}{0&b_1&1&0}{1}\nonumber\\
&-\EE{0&-1&1&1}{0&b_1&1&b_1}{1}+\EE{0&-1&1&1}{0&b_1&b_1&1}{1}-2 \EE{0&-1&1&1}{0&b_1&b_3&1}{1}\nonumber\\
&+2 \EE{0&-1&1&1}{0&b_1&b_3&b_1}{1}+\log\left(\frac{1}{u_2}\right)\times \nonumber\\
&\quad\left\{2 \EE{0,-1,1}{0,1,1}{1}+\EE{0,-1,1}{0,1,b_3}{1}+\EE{0,-1,1}{0,b_5,0}{1}+\EE{0,-1,1}{0,b_6,0}{1}\right\} \nonumber\\
&+\log\left(u_2\right)\Bigg\{\EE{0,-1,1}{0,0,1}{1}+\EE{0,-1,1}{0,0,b_3}{1}+2 \EE{0,-1,1}{0,1,0}{1}\nonumber\\
&\quad\quad\quad+\EE{0,-1,1}{0,b_5,b_3}{1}+\EE{0,-1,1}{0,b_6,b_3}{1}-2 \EE{0,-1,1}{0,b_1,b_3}{1}\Bigg\}+\log\left(\frac{1}{u_3}\right)\times\nonumber\\
& \left\{2 \EE{0,-1,1}{0,0,1}{1}-3 \EE{0,-1,1}{0,1,1}{1}+\EE{0,-1,1}{0,b_5,1}{1}+\EE{0,-1,1}{0,b_6,1}{1}\right\}\nonumber\\
& +\log\left(u_3\right)\Bigg\{-3 \EE{0,-1,1}{0,0,b_2}{1}+3 \EE{0,-1,1}{0,0,b_4}{1}+\EE{0,-1,1}{0,1,b_2}{1}\nonumber\\
&\quad\quad\quad-3\EE{0,-1,1}{0,1,b_4}{1}+\EE{0,-1,1}{0,b_5,b_2}{1}+\EE{0,-1,1}{0,b_6,b_2}{1}\Bigg\}\nonumber\\
& +\log\left(\frac{u_3}{u_2}\right)\left\{-2 \EE{0,-1,1}{0,1,b_1}{1}+\EE{0,-1,1}{0,b_5,b_1}{1}+\EE{0,-1,1}{0,b_6,b_1}{1}\right\}\nonumber\\
& +\EE{0,-1,1}{0,b_1,1}{1} \log\left(u_2 u_3\right)\nonumber\\
&+\EE{0,-1}{0,1}{1} \Bigg\{-2 \zeta_2+\Li_2\left(\frac{1}{u_2}\right)+\log^2\left(\frac{1}{u_2}\right)+\frac{1}{2} \log\left(\frac{1-u_2}{u_2}\right) \log\left(\frac{1}{u_2}\right)\nonumber\\
&\quad\quad\quad-2 \log \left(\frac{u_3}{u_2}\right)\log \left(\frac{1}{u_2}\right)+\frac{1}{2} i \pi  \log\left(\frac{1}{u_2}\right)+\frac{1}{2} \log\left(\frac{1-u_2}{u_2}\right) \log\left(\frac{1}{u_3}\right)\nonumber\\
&\quad\quad\quad+\frac{1}{2} i \pi  \log\left(\frac{1}{u_3}\right)+\frac{1}{2} \log\left(\frac{1-u_2}{u_2}\right) \log\left(\frac{u_3}{u_2}\right)+\frac{1}{2} i \pi  \log\left(\frac{u_3}{u_2}\right)\Bigg\}\nonumber\\
&+\EE{0,-1}{0,b_5}{1}\Bigg\{\zeta_2-\Li_2\left(\frac{1}{u_2}\right)-\frac{1}{2} \log^2\left(\frac{1}{u_2}\right)+\log\left(\frac{u_3}{1-u_2}\right) \log\left(\frac{1}{u_2}\right)\nonumber\\
&\quad\quad\quad-i \pi  \log\left(\frac{1}{u_2}\right)\Bigg\} +\EE{0,-1}{0,b_6}{1}\Bigg\{\zeta_2-\Li_2\left(\frac{1}{u_2}\right)-\frac{1}{2} \log^2\left(\frac{1}{u_2}\right)\nonumber\\
&\quad\quad\quad+\left(\log\left(\frac{u_3}{1-u_2}\right)-i \pi \right) \log\left(\frac{1}{u_2}\right)\Bigg\}\nonumber\\
& +2\EE{0,-1}{0,b_1}{1} \Bigg\{\Li_2\left(\frac{1}{u_2}\right)+\log
\left(\frac{u_2}{1-u_2}\right) \log \left(u_2\right)-i \pi \log \left(u_2\right)\Bigg\}\nonumber\\
& +\EE{0,-1}{0,0}{1}\Bigg\{(-\Li_2\left(\frac{1}{u_2}\right)-\log\left(\frac{u_2}{1-u_2}\right) \log \left(u_2\right)+i \pi \log \left(u_2\right)\Bigg\} \nonumber\\
&\EE{0}{0}{1} \Bigg\{\zeta_3+\Li_3\left(\frac{1}{u_2}\right)-\Li_3\left(-\frac{u_2}{1-u_2}\right)-\Li_2\left(\frac{1}{u_2}\right) \log\left(\frac{u_3}{u_2}\right)-\zeta_2 \log\left(\frac{u_2}{1-u_2}\right)\nonumber\\
&\quad\quad\quad+\log\left(u_2\right)\Big(\left(\log \left(\frac{1-u_2}{u_2}\right)+i \pi \right)\log \left(\frac{\left(y_2-1\right) y_3}{y_2 \left(y_3-1\right)}\right)\nonumber\\
&\quad\quad\quad\quad+\left(\log \left(\frac{1-u_2}{u_2}\right)+i\pi \right) \log\left(\frac{y_1 y_2-1}{y_1 y_3-1}\right)\Big)-\frac{1}{6}\log^3\left(\frac{u_2}{1-u_2}\right)\nonumber\\
&\quad\quad\quad+\log^2\left(u_2\right)\left(-\frac{1}{2} \log\left(\frac{u_2}{1-u_2}\right)+\frac{i \pi }{2}\right)\Bigg\}
\,,
\label{eq:pureresult}
\end{align}
}
where
\be
\begin{gathered}
b_1=\frac{1}{1-u_1}\,,\\
b_2=\frac{1}{u_3}\,,\\
b_3=\frac{\left(y_1 y_3-1\right) \left(y_1 y_2 y_3-1\right)}{\left(y_1-1\right) \left(y_1 y_2 y_3^2-1\right)}\,,\\
b_4=\frac{\left(y_1 y_3-1\right) \left(y_2 y_3-1\right)}{1-2 y_2 y_3+y_1 \left(y_3 y_2+y_2-2\right) y_3+y_3}\,,\\
b_5=\frac{1-y_1 y_3}{y_3-y_1 y_3}\,,\\
b_6=\frac{y_1 y_3-1}{y_1-1}\,.
\end{gathered}
\ee
We provide this expression in machine-readable format in an ancillary file, \texttt{ToyDoubleBox.m}.

We can then find an expression in terms of $\tilde{\Gamma}$ functions. This expression is even longer, as the integration kernels of the $\tilde{\Gamma}$ functions are not manifestly single-valued and thus at least two such kernels are needed to represent each kernel in the pure elliptic multiple polylogarithm formalism (see eq.~\ref{eq:kernelrelation}). Due to length we omit this expression here, and instead provide it in the ancillary file \texttt{ToyDoubleBox.m}.

Our result is pure once the relations in eqs.~\ref{eq:id1},~\ref{eq:id2}, and~\ref{eq:id3} are taken into account. This is as expected, both from the perspective of this integral's leading singularity and because it was described as a one-fold integral over pure polylogarithms. The form in eq.~\ref{eq:pureresult} is not manifestly symmetric in exchange of the cross-ratios, but this is also as expected because the functions depend on an explicit ordering of roots $\vec{a}$ which is not symmetric.

Our expression in terms of $\tilde{\Gamma}$ functions is also not manifestly symmetric. This is surprising, at least naively, because the definition of the functions seems to depend only on quantities that are symmetric in the cross-ratios, such as the elliptic modulus $\tau$. However, the existence of an integration endpoint at $\alpha=\infty$ does introduce an asymmetry into the problem: the leading term in the $\alpha\rightarrow\infty$ limit of $Q(\alpha)$ is proportional to $u_1 u_3$ and thus not manifestly symmetric, so the argument of these functions will not be either.

\section{Conclusions}\label{sec:conclusions}

We have constructed both pure eMPL and $\tilde{\Gamma}$ expressions for the toy model double-box. We expect this example to be useful to future investigations: much as six-particle kinematics has had an important role in investigations of polylogarithmic amplitudes (see for example refs.~\cite{Goncharov:2010jf,Dixon:2014xca,Basso:2015uxa,Caron-Huot:2020bkp,Basso:2020xts,Dixon:2021tdw,DelDuca:2022skz,Arkani-Hamed:2022rwr} and citations therein), being neither ``too simple'', nor ``too complex'', so we expect this toy model, depending on the same number of parameters, to be useful: more complex than the sunrise diagram, but simpler than the elliptic double-box for generic kinematics.

We noted that the symmetry of this diagram is not manifest in any of the forms we found. We expect that this symmetry can be made manifest using the symbol formalism developed for elliptic polylogarithms in ref.~\cite{Broedel:2018iwv}. It would be interesting to see if the diagram's symmetry is immediately manifest in this form, or if it requires identities between symbol letters to be made manifest, possibly requiring the symbol-prime formalism of ref.~\cite{Wilhelm:2022wow}.
\\\\

\noindent\textbf{Acknowledgements}

We thank Falko Dulat, Brenda Penante, Matthias Wilhelm, Cristian Vergu, and Chi Zhang for helpful discussions. This work was supported by the Danish National Research Foundation (DNRF91), the research grant 00015369 from Villum Fonden, and a Starting Grant (No. 757978) from the European Research Council.

\providecommand{\href}[2]{#2}\begingroup\raggedright\endgroup

\end{document}